\documentclass[12pt,letter]{article}
\usepackage{amsmath}
\usepackage{amssymb}
\usepackage[colorlinks=true,linkcolor=blue,urlcolor=blue]{hyperref}
\usepackage{epsf,graphicx,rotating,color}
\usepackage{bm,bbm}
\usepackage{float}
\newcommand{\eps}{\varepsilon}

\newcommand{\la}{\langle}
\newcommand{\ra}{\rangle}

\newcommand{\lam}{\lambda}

\newcommand{\rmd}{{\rm d}}
\newcommand{\rme}{{\rm e}}

\newcommand{\rmi}{{\rm i}}

\def\one{\mathbbm{1}}

\begin{document}

\title{A generalized fidelity amplitude for open systems}

\author{Thomas Gorin$^1$, H\' ector J. Moreno$^{2,3}$, Thomas H. Seligman$^{3,4}$\\
   \small\em $^1$ Departamento de F\'\i sica, Universidad de Guadalajara, 
     Guadalajara, Jal\'\i sco, M\' exico\\
   \small\em $^2$ Centro de Investigaci\' on en Ciencias, Universidad 
     Aut\' noma del Estado de Morelos, Cuernavaca,\\ 
   \small\em Morelos, M\' exico\\
   \small\em $^3$ Instituto de Ciencias F\'\i sicas, Universidad Nacional 
     Aut\' onoma de M\' exico, Cuernavaca,\\ 
   \small\em Morelos, M\' exico\\
   \small\em $^4$ Centro Internacional de Ciencias A. C., Cuernavaca, Morelos,
     M\' exico }

\date{}

\maketitle

\begin{abstract}
We consider a central system which is coupled via dephasing to an
open system, i.e. an intermediate system which in turn is coupled to another
environment. Considering intermediate and far environment as one composite 
system, the coherences in the central system are given in the form of fidelity
amplitudes for a certain perturbed echo dynamics in the composite environment.
On the basis of the Born-Markov approximation, we derive a master equation for 
the reduction of that dynamics to the intermediate system alone. In distinction 
to an earlier paper [arXiv: 1502.04143 (2015)] where we discussed the 
stabilizing effect of the far environment on the decoherence in the central 
system, we focus here on the possibility to use the measurable coherences in 
the central system for probing the open quantum dynamics in the intermediate 
system. We illustrate our results for the case of chaotic dynamics in the near 
environment, where we compare random matrix simulations with our analytical 
result.
\end{abstract}


\section{\label{I} Introduction}

Loss of fidelity and decoherence are the twin obstacles to successful 
applications of quantum information devices. Theoreticians like to consider the 
two separately, while in practical situations both will have destructive 
effects on quantum information flow.
In this spirit it is important to propose a functional definition for the
fidelity or better the fidelity amplitude of an open system. To achieve this we 
shall build on previous work. Almost 20 years ago a measurement of the fidelity 
amplitude in a quantum system was proposed~\cite{GCZ97}, though the word itself 
was not mentioned. A more detailed presentation for a kicked rotor followed 
years later~\cite{Haug05}, simultaneously with an extensive discussion of the 
theoretical framework~\cite{GPSS04}. The experiment for the kicked rotor was 
performed successfully in Ref.~\cite{WuToPr09}. The basic idea is that pure 
dephasing between a qubit and some system will cause the off diagonal elements 
or coherences of the density matrix of a pure superposition state of the qubit 
decay like the fidelity amplitude of the remaining system. Our proposition is 
to follow exactly the same reasoning in the case where the remaining system is 
open, and thus interpret the decaying coherences as a generalized fidelity 
amplitude of the intermediate system.

For this purpose we have to consider a situation discussed in a previous 
paper~\cite{MGS15}, where a central system coupled to two nested environments 
is analyzed to determine the effect of a far environment on the coherence of 
a central system not interacting directly with the far environment. The 
intermediate system or ``near environment'' would then be the open system we 
are analyzing using the qubit as a probe and as the source of the perturbation
causing the fidelity decay in the absence of the far environment.

Our paper will thus fulfill a double purpose: On the one hand we shall 
introduce a generalized fidelity amplitude and study its behaviour in a 
``quantum chaotic'' setting. On the other we shall deepen the understanding of 
the stabilizing effect on the central system of the coupling of the near 
environment to a far environment. Note that in a previous paper~\cite{MGS15} we 
introduced pure dephasing between the central system and the near environment, 
as a simplifying approximation to facilitate calculations, but for the present 
context this is an essential feature needed in order to establish the relation 
to fidelity decay of a closed system. 

Considering near and far environment as one closed system, the fidelity 
amplitude may be expressed as the expectation value of an echo 
operator~\cite{GPSZ06}, which describes the forward and backward evolution of 
an initial state with somewhat different Hamiltonians. In the present paper we 
shall derive a master equation which describes this echo dynamics reduced to 
the near environment alone. This master equation has the typical structure of
master equations of Kosakowski-Sudarshan-Lindblad 
form~\cite{GorKossSud76,Lin76} (henceforth referred to as Lindblad equation) 
and reduce to Lindblad form when the coupling to the central system becomes
zero. 

We then consider the random matrix model already used in Ref.~\cite{MGS15}. In 
that case, the master equation becomes very simple, and allows to obtain a 
closed integral equation for the generalized fidelity amplitude. We compare 
the solution of the integral equation with the generalized fidelity amplitude
obtained from numerical simulations, and thereby show that the new integral
equation applies for a very broad range of coupling strengths between near and
far environment. While the random matrix model used here, is much simpler than
the one used in Ref.~\cite{MGS15}, its effect on the fidelity amplitude is 
typically indistinguishable.

In the next section we shall fix notations and obtain the general master 
equation for describing the reduced echo dynamics in the near environment.
At the end of this section, we derive the master equation for the random
matrix model, which is largely equivalent to the model considered 
in~\cite{MGS15}. In Sec.~\ref{N} we compare the generalized fidelity amplitude 
obtained from the new integral equation with numerical simulations, and in the 
last section we draw conclusions.

\section{\label{M} Model}

In this section we describe the model in general. The full system consists of 
three parts, the central system, the near environment and a far environment. In 
the following section (Sec.~\ref{MH}) we start from a Hamiltonian formulation, 
and perform partial traces in order to obtain the reduced dynamics either in 
the central system or in the near environment. Assuming a dephasing coupling 
between central system and near environment, we can obtain the temporal 
evolution of the reduced density matrix in the central system, in terms of an 
asymmetric unitary evolution (perturbed echo dynamics) in the composite 
environment. In Sec.~\ref{MM} we then use the standard Born-Markov 
approximation to trace out the far environment, and thereby arrive at a master 
equation for the reduced echo dynamics in the near environment alone.

\subsection{\label{MH} Hamiltonian formulation}

For the coupling between central system and near environment, we assume that it
is of the dephasing type, i.e. that it is given by a single product term, where 
the factor acting in the central system commutes with the Hamiltonian 
describing the dynamics in the central system. Thus, the Hamiltonian for this 
part is given as
\begin{equation}
    H_{\rm c,e} = h_{\rm c} \otimes \one_\rme  +
    \one_{\rm c} \otimes H_{\rm e} + v_{\rm c} \otimes V_{\rm e} \; , 
\quad\text{where}\quad [h_{\rm c},v_{\rm c}]= 0 \; .
\label{MH:Hlambda}\end{equation}
This leads to the following expression for the time evolution of the reduced 
density matrix in the central system~\cite{GCZ97,GPSS04}
\begin{equation}
\varrho_{\rm c}(t)= {\rm tr}_\rme\big [\, \rme^{-\rmi H_{\rm c,e} t/\hbar}\,
   \varrho_{\rm c}(0)\otimes\varrho_\rme\, \rme^{\rmi H_{\rm c,e} t/\hbar}\, 
\big ] \; ,
\end{equation}
which yields for its individual matrix elements (coherences)
\begin{equation}
\varrho^{\rm c}_{jk}(t) = \varrho^{\rm c}_{jk}(0)\; 
   \rme^{-\rmi (\eps_j -\eps_k)/\hbar}\; f_\lambda(t) \; , \qquad
f_\lambda(t)= {\rm tr}\big [\, \rme^{-\rmi H_\lambda t/\hbar}\, \varrho_\rme\,
   \rme^{\rmi H_0 t/\hbar}\, \big ] \; .
\label{MH:fidamp0}\end{equation}
Here, we use the common eigenbasis of $h_{\rm c}$ and $v_{\rm c}$ to express
$\varrho_{\rm c}(t)$, and therefore the energies $\eps_j$ and $\eps_k$ are 
simply the corresponding eigenvalues of $h_{\rm c}$. In what follows we will 
focus on only one such matrix element, and therefore suppress the indices $j,k$ 
from now on. We thus set 
\begin{equation}
H_\lambda = H_\rme + \lambda\; V_\rme = H_\rme + \nu_j\, V_\rme
\quad\text{and}\quad
H_0= H_\rme + \nu_k\, V_\rme\; , 
\end{equation}
which implies that $\lambda = \nu_j - \nu_k$. This shows that under dephasing
coupling, the decoherence in the central system is given by the decay of the
fidelity or Lohschmidt echo in the near environment. Turning the argument the
other way around, this shows that it is possible to measure fidelity 
amplitudes, by coupling the system of interest (i.e. the near environment) to
a probe system, which at the same time provides the perturbation. Such 
experiments have been proposed and recently realized in different settings
using atom interferometry~\cite{GCZ97,Haug05,AndDav03,WuToPr09}.

\paragraph{Including the far environment} We now extend the model to include 
a far environment. We assume the far environment to be as simple as possible
and that it allows to be taken into account implicitly in the form of a 
quantum master equation. Hence, we write for the Hamiltonian of the full 
tripartite system:
\begin{equation}
H_{\lambda,\Gamma} = H_{\rm c,e}\otimes\one_{\rm f} 
   + \one_{\rm c}\otimes\; \big (\, \one_\rme\otimes H_{\rm f}
   + \gamma\; V'_\rme\otimes V_{\rm f}\, \big ) \; ,
\end{equation}
where $\Gamma = 2\pi\, N_{\rm e}\, \gamma^2/(\hbar d_{\rm f})$. Here, $V'_\rme$
and $V_{\rm f}$ are normalized in such a way that $\gamma^2$ gives the 
magnitude squared of a typical matrix element of the coupling term between near 
and far environment. With $d_{\rm f}$ being the average level spacing (i.e. the 
inverse level density) in the spectrum of $H_{\rm f}$, we see that $\Gamma$ is
just $N_\rme$ times the corresponding Fermi golden rule transition 
rate~\cite{fermi1974nuclear, Dirac243}. Finally, $\Gamma$ itself is related to 
the decoherence rate for a superposition of $N_\rme$ states in the near 
environment. In the simplest case (see Sec.~\ref{N}), $2\Gamma$ is precisely 
the decay rate of the purity in the intermediate system, in the case where the
coupling to the central system is set to zero.
%Within the analytical treatment and also in the numerical
%simulations, we will see that $\Gamma$ is the only parameter, needed to 
%quantify the effect of the far environment onto the echo-dynamics in the 
%near environment (aca decay of coherences in the central system).

Within the Hamiltonian model described by $H_{\lambda,\Gamma}$, the coupling 
to the far environment requires the following modification to the 
expression for the fidelity amplitude given in Eq.~(\ref{MH:fidamp0}):
\begin{equation}
f_{\lambda,\Gamma}(t)= {\rm tr}_{\rm e,f}\big [\, 
   \rme^{-\rmi H_{\lambda,\Gamma}\, t/\hbar}\; \varrho_{\rm e,f}(0)\; 
   \rme^{\rmi H_{0,\Gamma}\, t/\hbar}\, \big ] \; .
\label{MH:fidampgen}\end{equation}
Differing from the standard formalism, the unitary operators on the left and on 
the right hand side of the initial state are different. This is why the trace
may decrease in time, leading to the loss of coherence for superposition states
in the central system~\cite{MGS15}.

\subsection{\label{MM} Master equation for the echo dynamics}

We follow the standard derivation of the Born-Markov approximation, e.g. 
Sec.~3.3 of Ref.~\cite{BrePet02}. However, due to the asymmetric unitary
transformation implied in Eq.~(\ref{MH:fidampgen}), the following derivation 
requires some care. Let us denote the solution in the Hilbert space of near and 
far environment as
\begin{equation}
\varrho_{\rm e,f}(t)= \rme^{-\rmi H_{\lambda,\Gamma}\, t/\hbar}\; 
   \varrho_{\rm e,f}\; \rme^{\rmi H_{0,\Gamma}\, t/\hbar}
 = \rme^{-\rmi H_{\lambda,0}\, t/\hbar}\; X(t)\; \rme^{\rmi H_{0,0}\, t/\hbar} 
\; ,
\end{equation}
and thereby introduce $X(t)$ as the solution in the interaction picture
with respect to the coupling between near and far environment. From the von
Neumann equation for $\varrho_{\rm e,f}(t)$, 
\begin{equation}
\rmi\hbar\, \partial_t\, \varrho_{\rm e,f}(t) = H_{\lambda,\Gamma}\; 
   \varrho_{\rm e,f}(t) - \varrho_{\rm e,f}(t)\; H_{0,\Gamma} \; ,
\end{equation}
we obtain% a similar equation for $X(t)$:
\begin{equation}
\rmi\hbar\, \partial_t\, X(t) = \gamma\; \big [\,
   \tilde V_\lambda(t)\; X(t) - X(t)\; \tilde V_0(t)\, \big ] \; , \qquad 
\tilde V_\lambda(t)= \rme^{\rmi H_{\lambda,0}\, t/\hbar}\; 
   V_\rme'\otimes V_{\rm f}\; \rme^{\rmi H_{0,0}\, t/\hbar} \; .
\label{MM:Xevolve}\end{equation}
The aim of the Born-Markov approximation (to be worked out next) consists in
obtaining a master equation for the reduced dynamics in the near environment
alone. That is, an evolution equation for
$\varrho_\rme(t)= {\rm tr}_{\rm f}\big [ \varrho_{\rm e,f}(t)\, \big ]$ which 
may not be considered a real quantum state, since for the reasons discussed 
above, $\varrho_\rme(t)$ is neither Hermitian nor trace-preserving. This 
quasi-density operator is related to $X(t)$ as follows:
\begin{align}
\varrho_\rme(t) &= {\rm tr}_{\rm f}\big [ \rme^{-\rmi H_{\lambda,0} t/\hbar}\;
   X(t)\; \rme^{-\rmi H_{0,0} t/\hbar}\, \big ] 
 = {\rm tr}_{\rm f}\big [ U_\lambda(t)\otimes U_{\rm f}(t)\; X(t)\; 
      U_0(t)^\dagger\otimes U_{\rm f}(t)^\dagger\, \big ] \; , 
\end{align}
where $U_\lambda(t)= \rme^{-\rmi H_\lambda t/\hbar}$ and
$U_{\rm f}(t)= \rme^{-\rmi H_{\rm f} t/\hbar}$, since $H_{\lambda,0}$ and 
$H_{0,0}$ are both separable operators. Using the identities
\begin{align} 
&{\rm tr}_{\rm f}\big [ A\otimes\one\; X\, \big ] 
   = A\; {\rm tr}_{\rm f}[ X\, ] \; , \quad
 {\rm tr}_{\rm f}\big [ X\; B\otimes\one\, \big ] 
   = {\rm tr}_{\rm f}[ X\, ]\; B \notag\\
& {\rm tr}_{\rm f}\big [ \one\otimes A\; X\, \big ] 
   = {\rm tr}_{\rm f}\big [ X\; \one\otimes A\, \big ]\; , 
\end{align}
we find
\begin{equation}
\varrho_\rme(t)= U_\lambda(t)\; \tilde\varrho_\rme(t)\; U_0(t)^\dagger \; ,
\qquad \tilde\varrho_\rme(t)= {\rm tr}_{\rm f}\big [ X(t)\, \big ]\; .
\label{MM:rhoetilde}\end{equation}

\paragraph{Born-Markov approximation}
Now, we will formally integrate the differential equation for $X(t)$, 
Eq.~(\ref{MM:Xevolve}), and plug the result back into its right hand side:
\begin{align}
X(t) &= \varrho_{\rm e,f}(0) - \frac{\rmi\gamma}{\hbar}\int_0^t\rmd\tau\; 
   \big (\, \tilde V_\lambda(\tau)\, X(\tau) - X(\tau)\, \tilde V_0(\tau)\,
   \big ) \; , \notag\\
\rmi\hbar\, \partial_t\, X(t) &= \gamma\; \tilde V_\lambda(t)\;
   \Big [ \, \varrho_{\rm e,f}(0) - \frac{\rmi\gamma}{\hbar}\int_0^t\rmd\tau\; 
   \big (\, \tilde V_\lambda(\tau)\, X(\tau) - X(\tau)\, \tilde V_0(\tau)\,
   \big )\, \Big ] \notag\\
&\qquad - \gamma\; \Big [ \, \varrho_{\rm e,f}(0) 
  - \frac{\rmi\gamma}{\hbar}\int_0^t\rmd\tau\; \big (\, 
    \tilde V_\lambda(\tau)\, X(\tau) - X(\tau)\, \tilde V_0(\tau)\, \big )\, 
    \Big ]\; \tilde V_0(t) \notag\\
 &= \gamma\; \Big [ \, \tilde V_\lambda(t)\; \varrho_{\rm e,f}(0) - 
      \varrho_{\rm e,f}(0)\; \tilde V_0(t)\, \big ] 
  - \frac{\rmi\gamma^2}{\hbar}\int_0^t\rmd\tau\; \big [\, \tilde V_\lambda(t)\;
      \big (\, \tilde V_\lambda(\tau)\, X(\tau) - X(\tau)\, \tilde V_0(\tau)
      \, \big ) \notag\\
&\qquad\qquad\qquad - \big (\, \tilde V_\lambda(\tau)\, X(\tau) 
      - X(\tau)\, \tilde V_0(\tau)\, \big )\; \tilde V_0(t)\; \big ]
\end{align}
The next step consists in taking the partial trace with respect to the far
environment. In doing so, we will assume that the first term, which comes 
before the integral over $\tau$, will not contribute, i.e.
${\rm tr}_{\rm f}[\, \tilde V_\lambda(t)\; \varrho_{\rm e,f}(0) - 
      \varrho_{\rm e,f}(0)\; \tilde V_0(t)\, ] = 0$. While this is true in the
random matrix model, in general it might be necessary to take this term into
account. Even then it does not present any difficulty, since the term is known
before hand. With $\tilde\varrho_\rme(t) = {\rm tr}_{\rm f}[ X(t)\, ]$, we find
\begin{align}
\partial_t\, \tilde\varrho_\rme(t) &= -\, \frac{\gamma^2}{\hbar^2}
   \int_0^t\rmd\tau\; {\rm tr}_{\rm f}\big [ \tilde V_\lambda(t)\;
      \big (\, \tilde V_\lambda(\tau)\, X(\tau) - X(\tau)\, \tilde V_0(\tau)
      \, \big ) \notag\\
&\qquad\qquad\qquad - \big (\, \tilde V_\lambda(\tau)\, X(\tau) 
      - X(\tau)\, \tilde V_0(\tau)\, \big )\; \tilde V_0(t)\; \big ] \; .
\end{align}
Here, we perform the two crucial approximations: (i) the Born approximation,
which assumes that the influence of the near environment on the state of the
far environment is negligible: 
$X(\tau) \approx \tilde\varrho_\rme(\tau)\otimes \varrho_{\rm f}(0)$, and (ii) 
the Markov approximation, that the state of the near environment 
$\tilde\varrho_\rme(\tau)$ is changing slowly on the time scale of the
correlation function $\tilde V_\lambda(t-s)\, \tilde V_\lambda(t)$. We finally
assume that the $\tilde V_\lambda(t-s)\, \tilde V_\lambda(t)$ quickly 
approaches zero as $s$ increases, so that
\begin{align}
&\partial_t\, \tilde\varrho_\rme(t)= -\, \frac{\gamma^2}{\hbar^2}
   \int_0^t\rmd\tau\; {\rm tr}_{\rm f}\big [ \tilde V_\lambda(t)\;
      \big (\, \tilde V_\lambda(\tau)\, 
               \tilde\varrho_\rme(t)\otimes \varrho_{\rm f}(0) 
          - \tilde\varrho_\rme(t)\otimes \varrho_{\rm f}(0)\, \tilde V_0(\tau)
      \, \big ) \notag\\
&\qquad\qquad\qquad - \big (\, 
   \tilde V_\lambda(\tau)\, \tilde\varrho_\rme(t)\otimes \varrho_{\rm f}(0)
      - \tilde\varrho_\rme(t)\otimes \varrho_{\rm f}(0)\, 
            \tilde V_0(\tau)\, \big )\; \tilde V_0(t)\; \big ] \notag\\
&\qquad = -\, \frac{\gamma^2}{\hbar^2}
   \int_0^\infty\rmd s\; {\rm tr}_{\rm f}\big [ \tilde V_\lambda(t)\;
      \big (\, \tilde V_\lambda(t-s)\, 
               \tilde\varrho_\rme(t)\otimes \varrho_{\rm f}(0) 
          - \tilde\varrho_\rme(t)\otimes \varrho_{\rm f}(0)\, \tilde V_0(t-s)
      \, \big ) \notag\\
&\qquad\qquad\qquad - \big (\, 
   \tilde V_\lambda(t-s)\, \tilde\varrho_\rme(t)\otimes \varrho_{\rm f}(0)
      - \tilde\varrho_\rme(t)\otimes \varrho_{\rm f}(0)\, 
            \tilde V_0(t-s)\, \big )\; \tilde V_0(t)\; \big ]  \; .
\end{align}
This equation is the equivalent of the so called Redfield 
equation~\cite{Redfield57}. Next, we will consider each term separately and 
take advantage of the fact that $\tilde V_\lambda(t)$ is a tensor product 
operator with respect to the near and the far environment:
\[ \tilde V_\lambda(t)= U_\lambda(t)^\dagger\; V_\rme'\, U_\lambda(t)
      \otimes U_{\rm f}(t)^\dagger\; V_{\rm f}\; U_{\rm f}(t) \; , \]
such that due to ${\rm tr}_{\rm f}[ A\otimes B] = A\; {\rm tr}_{\rm f}[B]$. It 
is now natural to assume $\varrho_{\rm f}(0)$ to be diagonal in the eigenbasis 
of $H_{\rm f}$, i.e. $[\varrho_{\rm f}(0), H_{\rm f}] = 0$ (such would be the 
case for a thermal state). Then we find
\begin{align}
&\partial_t\, \tilde\varrho_\rme(t)= -\, \frac{\gamma^2}{\hbar^2}
   \int_0^\infty\rmd s\; C(s)\; \big\{\, 
   U_\lambda(t)^\dagger\, V_\rme'\, U_\lambda(s)\, V_\rme'\, U_\lambda(t-s)\; 
   \tilde\varrho_\rme(t) \notag\\
&\qquad\qquad\qquad -\;
   U_\lambda(t)^\dagger\, V_\rme'\, U_\lambda(t)\, \tilde\varrho_\rme(t)\,
   U_0(t-s)^\dagger\, V_\rme'\, U_0(t-s) \notag\\
&\qquad\qquad\qquad -\;
   U_\lambda(t-s)^\dagger\, V_\rme'\, U_\lambda(t-s)\, \tilde\varrho_\rme(t)\,
   U_0(t)^\dagger\, V_\rme'\, U_0(t) \notag\\
&\qquad\qquad\qquad +\;
   \tilde\varrho_\rme(t)\, U_0(t-s)^\dagger\, V_\rme'\, U_\lambda(s)^\dagger\, 
   V_\rme'\, U_0(t)\, \big \} \; ,
\end{align}
with the real function 
$C(s)= {\rm tr}_{\rm f}\big [ U_{\rm f}(s)^\dagger\, V_{\rm f}\, 
      U_{\rm f}(s)\, V_{\rm f}\, \varrho_{\rm f}\, \big ]$.
Going back to the Schr\" odinger picture, we obtain
\begin{align}
&\rmi\hbar\, \partial_t\, \varrho_\rme(t) = H_\lambda\, \varrho_\rme(t)
   - \varrho_\rme(t)\, H_0 \notag\\
&\qquad - \frac{\gamma^2}{\hbar^2}\; \big\{ \,
   V_\rme'\, \Gamma_\lambda\, \varrho_\rme(t) - V_\rme'\, \varrho_\rme(t)\,
   \Gamma_0 - \Gamma_\lambda\, \varrho_\rme(t)\, V_\rme'  + \varrho_\rme(t)\,
   \Gamma_0\, V_\rme' \, \big\} \; ,
\label{MM:MasterEqGen}\end{align}
where
\begin{equation}
\Gamma_\lambda = \int_0^\infty\rmd s\; C(s)\; U_\lambda(s)\, V_\rme'\, 
   U_\lambda(s)^\dagger \; .
\end{equation}
In the simplest case, $C(s)$ may be approximated by a delta function, such
that $\Gamma_\lambda = C_0\, V_\rme'/2$, where $C_0$ is the area under the
function $C(s)$, and the factor $1/2$ comes from the fact that the integral
goes only over the positive half axis. In this case, the master 
equation~(\ref{MM:MasterEqGen}) is of Lindblad form with the Hermitian Lindblad
operator $\sqrt{C_0}\, V_\rme'$. Note however, that there are many interesting
cases, where $C(s)$ has a different functional behavior, which then yields a
rather unusual dissipation term. 

\paragraph{RMT model} We here use the approach of constructing the average 
density matrix proposed in~\cite{GPKS08} rather then calculating properties for 
each member of the ensemble and averaging these properties afterwards. For 
details see~\cite{PGS07,Kap07}. Let us assume the master equation derived above 
is of Lindblad form with a single Lindblad operator $V_\rme'$. Then, we may 
choose $C_0 = 1$ without restriction. Let us further assume that $V_\rme'$ is a 
fixed member of the Gaussian orthogonal (GOE) or unitary ensemble (GUE). This 
may be justified for an environment dominated by chaotic dynamics or disorder. 
The Born-Markov approximation used above implies some coarse-graining in time, 
i.e.  averaging over a small time interval. Assuming ergodicity, this averaging 
may be replaced by averaging over the random matrix ensemble. In the case of 
the Gaussian ensembles, it is then enough to use the fact that 
$\la V_\rme'{}_{ij}\, V_\rme'{}_{kl}\ra = \delta_{jk}\, \delta_{il}$ (GUE) or
$\la V_\rme'{}_{ij}\, V_\rme'{}_{kl}\ra = \delta_{jk}\, \delta_{il} +
   \delta_{ik}\, \delta_{jl}$ (GOE), to arrive at the following very simple 
master equation:
\begin{equation}
\rmi\hbar\, \partial_t\, \varrho_\rme(t) = H_\lambda\, \varrho_\rme(t)
   - \varrho_\rme(t)\, H_0 - \Gamma\; \big\{ \, \varrho_\rme(t) 
   - \frac{1}{N_\rme}\; {\rm tr}\, \varrho_\rme(t)\, \big \} \; .
\label{MM:RMTmastereq}\end{equation}
The dissipation term of this equation had been obtained previously for a random 
matrix model where the coupling matrix is a full random matrix in the product 
Hilbert space of near and far environment~\cite{GPKS08}. Note that 
Eq.~(\ref{MM:rhoetilde}) implies
\begin{equation}
f_{\lambda,\Gamma}(t) = {\rm tr}\, \varrho_\rme(t)
   = {\rm tr}_\rme\big [\, M_\lambda(t)\, \tilde\varrho_\rme(t)\, \big ] \; ,
\qquad
M_\lambda(t)= U_0(t)^\dagger\, U_\lambda(t) \; .
\end{equation}

\subsection{\label{ME} Evolution equation for the generalized fidelity 
   amplitude} 

To solve the master equation~(\ref{MM:RMTmastereq}), we return to the 
interaction picture and separate off the exponential decay from the solution:
\begin{equation}
 \tilde\varrho_\rme(t) = \rme^{-\Gamma t}\; \Omega(t) \quad :\quad 
   \partial_t\, \Omega(t) = \frac{\Gamma}{N_\rme}\; M_\lambda^\dagger(t)\; 
   {\rm tr}_\rme\big [\, M_\lambda(t)\, \Omega(t)\, \big ] \; .
\end{equation}
We now consider the corresponding integral equation 
\begin{equation}
\Omega(t) = \Omega(0) + \frac{\Gamma}{N_\rme}\int_0^t\rmd\tau\; 
   M_\lambda^\dagger(\tau)\; {\rm tr}_\rme\big [\, M_\lambda(\tau)\, 
   \Omega(\tau)\, \big ] \; ,
\end{equation}
and define
\begin{equation}
\phi_{\lambda,\Gamma}(t) 
   = {\rm tr}_\rme\big [\, M_\lambda(t)\, \Omega(t)\, \big ]
 \quad\text{such that}\quad 
f_{\lambda,\Gamma}(t) = \rme^{-\Gamma t}\; \phi_{\lambda,\Gamma}(t) \; .
\end{equation}
Eventually, we are only interested in the generalized fidelity amplitude 
$f_{\lambda,\Gamma}(t)$. It is then convenient to derive an evolution equation 
for $\phi_{\lambda,\Gamma}(t)$ directly. To this end, we multiply the above 
integral equation with $M_\lambda(t)$, and take the trace. This yields
\begin{equation}
\phi_{\lambda,\Gamma}(t) = {\rm tr}_\rme\big [\, M_\lambda(t)\, 
   \varrho_\rme(0)\, \big ] + \frac{\Gamma}{N_\rme}\int_0^t\rmd\tau\;
   {\rm tr}_\rme\big [\, M_\lambda(t-\tau)\, \big ]\;
   \phi_{\lambda,\Gamma}(\tau)
\end{equation}
The first term of the RHS is just the fidelity amplitude in the near 
environment without coupling to the far environment. We will denote this 
function as $f_\lambda(t)= f_{\lambda,0}(t)$. The term 
$N_\rme^{-1}\, {\rm tr}_\rme\big [\, M_\lambda(t-\tau)\, \big ]$ is the same
type of fidelity amplitude, just that here the initial state is the maximally
mixed state $\one/N_\rme$. That function we will denote with 
$\bar f_\lambda(t)$. In the numerical simulations in Sec.~\ref{N} we will 
assume that $\varrho_\rme(0)$ is the same maximally mixed state, such that
$f_\lambda(t)= \bar f_\lambda(t)$. Thus, we may write
\begin{equation}
\phi_{\lambda,\Gamma}(t) = f_\lambda(t) + \Gamma\int_0^t\rmd\tau\;
   \bar f_\lambda(t-\tau)\; \phi_{\lambda,\Gamma}(\tau) \; .
\label{ME:genfidampInteq}\end{equation}
Since the integral in this expression is precisely equal to the convolution of
the functions $\bar f_\lambda(t)$ and $\phi_{\lambda,\Gamma}(t)$, the equation
can be solved formally by a Laplace transformation~\cite{Arfken05}. It leads to
\begin{equation}
\phi_{\lambda,\Gamma}(t) = \frac{1}{2\pi\rmi}\lim_{T\to\infty}\int_{-T}^T
   \rmd s\; \frac{\rme^{st}\, F(s)}{1-\Gamma\, \bar F(s)}\; , \qquad
F(s)= \int_0^\infty\rmd t\; \rme^{-st}\; f_\lambda(t) \; ,
\end{equation}
and similarly for $\bar F(s)$. In this expression, the ratio
$F(s)\, [1-\Gamma\, \bar F(s)]^{-1}$ can be formally expanded into a power
series in $\Gamma$, where the inverse Laplace transform of powers of $F(s)$ and
$\bar F(s)$ yield iterated convolutions of the original fidelity amplitudes.
Terminating the series at first order in $\Gamma$ leads to the approximate
expression for $f_{\lambda,\Gamma}(t)$ derived in~\cite{MGS15}.  

In practice, it is more convenient to solve the integral equation directly
using a numerical equal-stepsize integration scheme. The theoretical curves
compared to random matrix calculations in the following section, are obtained
via the trapezoidal rule. We noticed that the larger $\Gamma$ the smaller the
required stepsize, which does not allow a straight forward exploration of the
large $\Gamma$ limit.

\section{\label{N} Numerical results}

In this section, we verify the integral equation~(\ref{ME:genfidampInteq}) for
the generalized fidelity amplitude, when the coupling to the far environment
can be described by an unitarily invariant dissipation term, see
Eq.~(\ref{MM:RMTmastereq}). We have shown that this situation occurs when the
dynamics in the near environment can be described by a random matrix ensemble,
for instance in the case when the quantum chaos conjecture applies.

For the numerical simulations, we use the methodology of Ref.~\cite{MGS15}, 
writing the evolution equation for $\varrho_\rme(t)$ in super-vector and 
super-matrix form, and solving the resulting system of differential equations
by diagonalization. Since the number of equations is of order $N_\rme^2$, we 
are restricted to relatively small dimensions, $N_\rme = 50$. For each case,
we perform three independent numerical averages over $n_{\rm run} = 1000$ 
realizations. This allows to estimate (roughly) the statistical uncertainty 
of the numerical results. 

To obtain a solution of the integral equation, we employ the method explained
at the end of the previous section, which is, as far as precision and computer
workload are concerned, equivalent to direct numerical integration. Hence, for
not to large values of $\Gamma$, we can obtain highly accurate results in very
short computation times. For the model considered, the fidelity amplitude 
$f_\lambda(t)$ without far environment, would be given by the universal random 
matrix result first derived in Ref.~\cite{StoSch04b}. However, for simplicity 
we use the exponentiated linear response formula from Ref.~\cite{GPS04}.

\begin{figure}
     \includegraphics[width=0.8\textwidth]{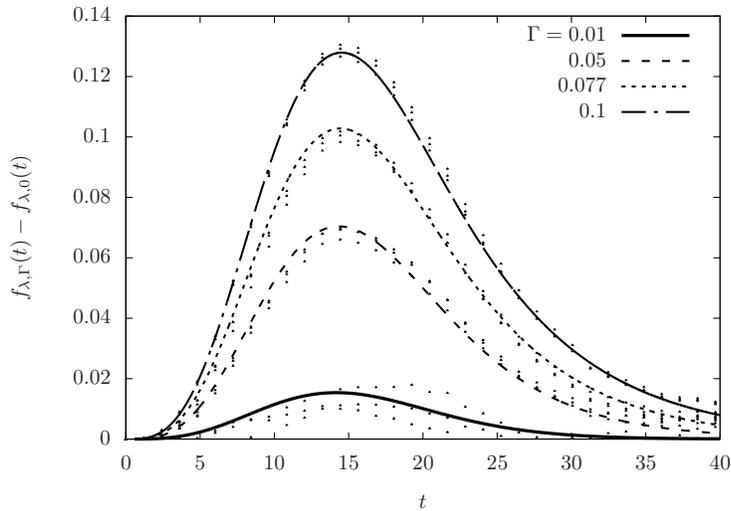}
 \caption{Generalized fidelity amplitude $f_{\lam,\Gamma}(t)$ subtracted by 
   $f_{\lam}(t)$ for $\lambda=0.1$ and different values of $\Gamma$: 
   $\Gamma= 0.01$ (solid line), $0.05$ (dashed line), $0.077$ (dotted line), 
   and $0.1$ (dashed-dotted line).
   The lines show the solutions of Eq.(\ref{ME:genfidampInteq}) and the 
   points show the corresponding numerical simulations.}
    \label{fig2-n50-l05}
\end{figure}

\begin{figure}
 \includegraphics[width=0.8\textwidth]{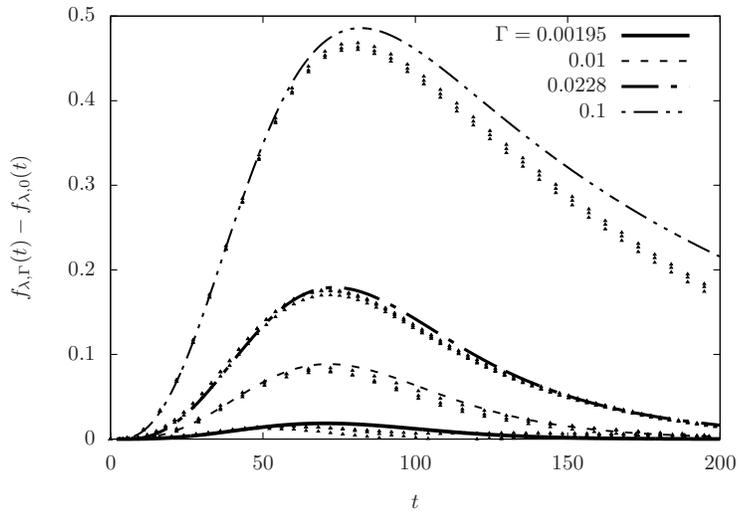}
 \caption{Generalized fidelity amplitude $f_{\lam,\Gamma}(t)$ subtracted by 
   $f_{\lam}(t)$ for $\lambda=0.02$ and different values of 
   $\Gamma$: $\Gamma= 0.00195$ (solid line)
   $0.01$ (dashed line) $0.02285$ (dashed-dotted line) and $0.1$ 
   (dashed-dashed-dotted-dotted line).
   The dashed lines show the solutions of Eq.(\ref{ME:genfidampInteq}) and the 
   points show the corresponding numerical simulations.}
    \label{fig3-n50-l01}
\end{figure}

Below, we consider two different strengths of the dephasing coupling. In 
Fig.~\ref{fig2-n50-l05} this is $\lambda = 0.1$ which corresponds to the 
cross-over regime where the decay of $f_\lambda(t)$ is intermediate between 
exponential and Gaussian decay. In Fig.~\ref{fig3-n50-l01}, $\lambda= 0.02$ 
which corresponds to the perturbative regime where the decay of $f_\lambda(t)$ 
is dominantly Gaussian. In both figures, the time scales are chosen such that
the Heisenberg time is at $t=2\pi$.

In order to compare the theoretical prediction with the numerical simulations
on a finer scale, we will plot the difference between the generalized 
fidelity amplitude $f_{\lambda,\Gamma}(t)$ and the fidelity amplitude 
$f_\lambda(t)= f_{\lambda,0}(t)$ without coupling to the far environment. 
%For the numerical simulations, both functions are obtained by random matrix 
%simulations, whereas for the theoretical prediction, we use the exponentiated
%linear response approximation. 

In Fig.~\ref{fig2-n50-l05} we can clearly see the influence of the coupling to
the far environment, on the coherence measured in the central system. The 
effect scales with $\alpha= \Gamma/\lambda$, which, in this figure ranges from
$\alpha= 0.1$ to $\alpha= 1$. Since the difference plotted is positive and
increasing with $\alpha$, we confirm the effect discussed in Ref.~\cite{MGS15},
that increasing the coupling to the far environment, stabilizes the coherence
in the central system. We also observe, that the theory obtained from 
Eq.~(\ref{ME:genfidampInteq}) agrees very well with the numerical simulations 
for all values of $\alpha$ and for all times. 

In Fig.~\ref{fig3-n50-l01} we repeat the comparison but for the case of
$\lam=0.02$, which is well in the perturbative regime, where $f_\lambda(t)$ 
shows a Gaussian decay. Again we compare numerical simulations with the 
theoretical prediction from Eq.~(\ref{ME:genfidampInteq}), for different values
of $\alpha$. Here, these values range from $\alpha = 0.0975$ up to 
$\alpha= 5$. In this range, the stabilizing effect of the coupling to the far
environment is even stronger. However, we also observe that the theoretical 
prediction systematically overestimates the effect, the larger $\alpha$ the 
larger the deviation. We suspect that these deviations might be a finite 
$N_\rme$ effect.

\section{Conclusions}

We have presented a generalized fidelity amplitude, which starts out from the
only direct measurement scheme for the fidelity amplitude~\cite{GCZ97}, which 
uses a central system as a probe for the echo dynamics in the environment.
We open the system to be probed to an additional outer environment and consider 
the decay of coherences of the probe. This we define to be proportional to the 
generalized fidelity amplitude. While this ensures that the quantity is 
measurable, it remains to be investigated how it relates to other 
generalizations of fidelity to open systems, such as the one by 
Josza~\cite{Josza}.

We obtain a  more general master equation, Eq.~(\ref{MM:MasterEqGen}), which 
may be applied to integrable and non-integrable models for open quantum 
systems. The random matrix model considered here to illustrate our results, is 
slightly different from the model considered in~\cite{MGS15}, since there we 
considered the far environment to be a bath of harmonic oscillators described 
by the Caldeira-Leggett model of random Brownian motion~\cite{CalLeg83}. 
Nevertheless, its effect on the near environment is equivalent and can be 
described by the same unique parameter $\Gamma$. We derived a new exact 
integral equation for the generalized fidelity amplitude, which constitutes 
significant progress comparing to previous work. Expanding this equation to 
lowest order in $\Gamma$ leads to the approximate analytical formula obtained 
in~\cite{MGS15}. In Sec.~\ref{N}, we find that its solution agrees with 
numerical simulations over a broad range of coupling strengths. This confirms 
the general experience, that increasing the coupling between near and far 
environment protects the coherence in the central system.

\paragraph{Acknowledgments} We thank C. Pineda and H.-J. St\"ockmann for
helpful discussions, and acknowledge the hospitality of the Centro 
Internacional de Ciencias, where these discussions took place. Finally, we
acknowledge financial support from CONACyT through the grants CB-2009/129309 
and 154586 as well as UNAM/DGAPA/PAPIIT IG 101113.

\end{document}